\documentclass[12pt]{article}
\usepackage{amsmath,amssymb,amscd,cite}
\newtheorem{thm}{Theorem}[section]
\newtheorem{prop}[thm]{Proposition}
\newtheorem{lem}[thm]{Lemma}

\newtheorem{defi}[thm]{Definition}
\newcommand{\pf}{{\bf Proof. \ }}
\newcommand{\qed}{\hfill $\Box$ \\}
\newtheorem{rem}[thm]{Remark}

\date{}
\begin{document}
\title{Simplex and MacDonald Codes over $R_{q}$}
\author{K. Chatouh, K. Guenda, T. A. Gulliver and L. Noui}
\date{21-05-2015}
\maketitle

\begin{abstract}
In this paper, we introduce the homogeneous weight and
homogeneous Gray map over the ring
$R_{q}=\mathbb{F}_{2}[u_{1},u_{2},\ldots,u_{q}]/\left\langle u_{i}^{2}=0,u_{i}u_{j}=u_{j}u_{i}\right\rangle$ for $q \geq 1$.
We also consider the construction of simplex and MacDonald codes of types $\alpha$ and $\beta$ over this ring.
Further, we study the properties of these codes such as their binary images and covering radius.
\end{abstract}


\section{ Introduction}
Codes over rings have been of significant research interest since the pioneering work of Hammons et al. \cite{Hammons}
on codes over $\mathbb{Z}_{4}$.
Many of their results have been extended to finite chain rings such as Galois rings and
rings of the form $\mathbb{F}_{2}[u]/\left\langle u^{m}\right\rangle$.
Recently, as a generalization of previous studies~\cite{yildiz1,yildiz2}, Dougherty et al.~\cite{dougherty} considered codes over
an infinite class of rings, denoted $R_q$.
These rings are finite and commutative, but are not finite chain rings.
Motivated by the importance of the simplex and MacDonald codes which have been defined over several finite commutative rings~\cite{alashkar,guptaGulliv,gupta},
in this work, we define the homogeneous weight over $R_q$ and present simplex codes and
MacDonald codes over this ring.
The properties of these codes are studied, particularly the weight enumerators and covering radius.
Further, the binary images of these codes are considered.

The remainder of this paper is organized as follows.
In Section 2, some preliminary results are given concerning the ring $R_q$ and codes over this ring.
Further, we define the homogeneous weight and its Gray map.
The simplex codes of type $\alpha$ and their properties and binary images
are given in Section 3, while
the simplex codes of type $\beta$ and their properties and binary images
are given in Section 4.
In Section 5, the MacDonald codes of types $\alpha$ and $\beta$ are presented along with their binary images.
Section 6 presents the repetition codes and considers some properties of these codes, in particular the covering radius.
Finally, in Section 7 the covering radius of the Simplex and MacDonald codes of types $\alpha$ and $\beta$ are studied.

\section{Preliminaries}
Let $R$ be a finite commutative ring and $R^{n}$ the set of all $n$-tuples over $R$.
Hence, $R^{n}$ is an $R$-module.
A code $C$ of length $n$ over $R$ is a non-empty subset of $R^{n}$.
A submodule $C$ of $R^{n}$ is called a linear code, and a code $C$ is called free if it is a free $R$-module.
Let $|C|$ denote the cardinality of $C$.
If $\left\vert C\right\vert = M$, then $C$ is called an $(n,M)$ code.
For any two vectors (or codewords)
$x=\left( x_{1},x_{2},\ldots,x_{n}\right),
y=\left( y_{1},y_{2},\ldots,y_{n}\right) \in R^{n}$,
the inner product is defined as
\begin{equation*}
\left\langle x,y\right\rangle =\sum\limits_{i=1}^{n}x_{i}y_{i}\in R.
\end{equation*}
Let $C\subseteq R^{n}$ be a code of length $n$ over $R$.
The dual code of $C$ is defined as
\begin{equation*}
C^{\perp }=\left\{ x\left\vert \left\langle x,y\right\rangle =0,\text{for
all }c\in C\right. \right\}.
\end{equation*}
%The Lee distance between the vector $x$ and $y\in R^{n}$is defined as%
%\begin{equation*}
%d\left( x,y\right) =\sum_{i=1}^{n}w\left( x_{i}-y_{i}\right).
%\end{equation*}
%The quality $d_{\min }=\left\{ d\left( x,y\right) /x,y\in C,x\neq y\right\} $
%is called the minimum distance of $C$.

%\subsection{The Ring $R_{q}=\mathbb{F}_{2}[u_{1},u_{2},...,u_{q}]/\left%
%\langle u_{i}^{2}=0,u_{i}u_{j}=u_{j}u_{i}\right\rangle $}
%
Let $q\geq 2$ be a positive integer.
Then the ring
$R_{q}=\mathbb{F}_{2}[u_{1},u_{2}, \ldots,u_{q}]/
\left\langle u_{i}^{2}=0,u_{i}u_{j}=u_{j}u_{i}\right\rangle$
is given recursively by
\begin{equation*}
R_{q}=\mathbb{F}_{2}[u_{1},u_{2}, \cdots, u_{q}] /\left\langle
u_{i}^{2}=0,u_{i}u_{j}=u_{j}u_{i}\right\rangle = R_{q-1}+u_{q}R_{q-1}.
\end{equation*}
For every subset $A\subseteq \left\{ 1,2, \cdots,q\right\}$ we have
\begin{equation*}
u_{A}=\prod_{i\in A}u_{i},
\end{equation*}
with the convention that $u_{\emptyset }=1$.
Then all elements of $R_{q}$ can be expressed by
\begin{equation*}
\sum_{A\subseteq \left\{ 1,2,\ldots ,q\right\} }c_{A}u_{A},\text{with }c_{A}\in
\mathbb{F}_{2}.
\end{equation*}
The following lemmas proved by Dougherty et al.~\cite{dougherty}
gives some important properties of $R_q$.
\begin{lem}
The ring $R_{q}$ is a local commutative ring with $\left\vert
R_{q}\right\vert =2^{2^q}$.
The unique maximal ideal $m_q$ consists of all non-units and
$\left\vert m_{q}\right\vert =\frac{\left\vert R_{q}\right\vert }{2}$.
\end{lem}
\begin{prop}
\label{lem:33} \hfill
\begin{itemize}
\item[(i)] For any $a\in R_q$, we have
\[
\begin{array}{lll}
a\cdot (u_{1}u_{2} \cdots u_{q}) & = & \left\{
\begin{array}{ll}
0 & \mbox{if } a \mbox{ is a non-unit}, \\
u_{1}u_{2}\cdots u_{q} & \mbox{if } a \mbox{ is a unit}.
\end{array}
\right.
\end{array}
\]
\item[(ii)] For any unit $a\in R_q$ and $x\in R_q$, we have
\[a \cdot x=u_{1}u_{2} \cdots u_{q} \Leftrightarrow x=u_{1}u_{2} \cdots u_{q}.
\]
We denote the set of units of $R_q$ by $\mathfrak{U}(R_q)$ and non-units by $\mathfrak{D}(R_q)$.
It is clear that
\[
\vert \mathfrak{U}(R_q) \vert=\vert \mathfrak{D}(R_q) \vert=2^{2^{q}-1} \mbox{ and } \mathfrak{U}(R_q)= \mathfrak{D}(R_q)+1.
\]
A linear code of length $n$ over $R_q$ is defined to be an $R_q$-submodule of $R_q^{n}$.
\end{itemize}
\end{prop}

\subsection{The Lee and Homogeneous Weights over $R_{q}$ and the Gray Maps}
\subsubsection{The Lee Weight over $R_{q}$ and the Gray Map}
Let the order on the subsets of $\left\{ 1,2,\ldots,q\right\}$ be
\begin{equation*}
\left\{ 1,2,\ldots,q\right\} =\left\{ 1,2,\ldots,q-1\right\} \cup \left\{q\right\}.
\end{equation*}
With this order, the Gray map is defined as follows
\begin{equation*}
\begin{array}{ccccc}
\Psi _{Lee} & : & R_{q} & \rightarrow & \mathbb{F}_{2}^{2^{^{q}}},
\end{array}
\end{equation*}
with
\begin{equation*}
\Psi _{Lee}\left( u_{A}\right) =\left( c_{B}\right) _{B\subset \left\{
1,2,\ldots,q\right\} },
\end{equation*}
and
\begin{equation*}
c_{B}=\left\{
\begin{array}{cc}
1 & \text{ \ \ if } B\subset A, \\
0 & \text{ \ \ otherwise.}
\end{array}
\right.
\end{equation*}
We can extend $\Psi _{Lee}$ to all elements of $R_{q}$ and define the Lee weight of an element in $R_{q}$ as
the Hamming weight of its image.
This is a linear distance preserving map from $R_{q}^{n}$ to $\mathbb{F}_{2}^{{2^q }n}$.
It follows immediately that
\begin{equation*}
w_{Lee}(u_{A})=2^{\left\vert A\right\vert }.
\end{equation*}
Hence we have the following lemma.
\begin{lem}
If $C$ is a linear code over $R_q$ of length $n$, cardinality $2^k$ and minimum Lee weight $d_{Lee}$,
then $\Psi_{Lee}(C)$ is a binary linear code with parameters $[2^{2^{q}}n,k,d_{Lee}]$.
\end{lem}

\subsubsection{The Homogeneous Weight over $R_{q}$ and the Gray Map}

Several weights can be defined over rings.
A weight on a code $C$ over the ring $R_{q}$ is called homogeneous if it satisfies the following assertions.
\begin{defi}
\cite[p. 19]{Greferath}
A real valued function $w$ on the finite ring $R_{q}$ is called
a (left) homogeneous weight if $w(0) = 0$ and the following are true.
\begin{enumerate}
\item[(i)] For all $x, y \in R_{q}$, $R_{q}x = R_{q}y$ implies $w(x) = w(y)$.
\item[(ii)] There exists a real number $\eta$ such that
\[ \sum_{y\in R_{x}} w(y)=\eta\vert R_{q}x \vert \  for\  all\  x\in R_{q}-\{0 \}.  \]
\end{enumerate}
\end{defi}
The number $\eta$ is the average value of $w$ on $R_{q}$, and from condition $(i)$ we can
deduce that $\eta$ is constant on every non-zero principal ideal of $R_{q}$.

Honold \cite{Honold} described the homogeneous weight on $R_q$ in terms of
generating characters.

\begin{prop}
\cite{Honold}
\label{prop:44} Let $R_{q}$ be a finite ring with generating character $\chi$.
Then every homogeneous weight on $R_{q}$ is of the form
\[
\begin{array}{lllll}
w & : & R_{q} & \rightarrow & \mathbb{R} \\
&  & x & \mapsto & \gamma\left[ 1-\frac{1}{\left\vert R_{q}^{\times}\right\vert }\sum_{u\in R_{q}^{\times }}\chi (xu)\right]
\end{array}
\].
\end{prop}

The homogeneous weight on $R_q$ will be obtained using Proposition \ref{prop:44}.
Recall from \cite{dougherty} that the following is a generating character for the ring $R_{q}$
\[
\chi\left(\sum\limits
_{\substack{ A\subseteq \left\{ 1,2,\ldots,q\right\}  \\ c_{\varnothing }=0\vee A \neq
\varnothing }}c_{A}u_{A} \right)=(-1)^{wt(c)},
\]
where $wt(c)$, denotes the Hamming weight of the $\mathbb{F}_{2}$-coordinate vector of the
element in the basis $\{u_{A};A \subseteq \{1,2,\cdots,q\}\}$.
We then have
\[
\begin{array}{l}
\chi(0)=1\\
\chi(1)=\chi(u_{1})=\cdots=\chi(u_{q})=\chi(u_{1}u_{2})=\cdots=\chi(u_{1}u_{2}\cdots u_{q})=-1\\
\chi(1+u_{1})=\chi(u_{1}+u_{2})=\cdots=\chi(u_{q}+u_{1}u_{2}\cdots u_{q})=1\\
\chi(1+u_{1}+u_{2})=\chi(u_{1}+u_{2}+u_{3})=\cdots=\chi(u_{q-1}+u_{q}+u_{1}u_{2}\cdots u_{q})=-1\\
\vdots\\
\chi\left(1+\sum\limits_{\substack{ A\subseteq \left\{ 1,2,\ldots,q\right\}  \\ c_{\varnothing }=0\vee A \neq
\varnothing }}c_{A}u_{A} \right)=1.
\end{array}
\]

The following Lemma from \cite[Theorem 2]{Honold} will be key in proving the main theorem concerning the homogeneous
weight on $R_{q}$.
\begin{lem}
\label{lem:45}
Let $x$ be an element in $R_{q}$ such that $x \neq 0$ and $x \neq u_{1}u_{2}\cdots u_{q}$.
Then
\[
\sum_{a\in R_{q}} \chi(a \cdot x)=0.
\]
\end{lem}
\begin{thm}
The homogeneous weight on $R_{q}$ is
\[
\begin{array}{lll}
w_{hom}(x) & = & \left\{
\begin{array}{ll}
0 & \mbox{if } x=0, \\
2\gamma & \mbox{if } x=u_{1}u_{2}\cdots u_{q},\\
\gamma & \mbox{otherwise}.
\end{array}
\right.
\end{array}
\]
\end{thm}

\pf
Let $x=u_{1}u_{2}\cdots u_{q}$.
Then by Proposition \ref{lem:33}, $a\cdot x=x$ for all $a\in \mathfrak{U}(R_{q})$,
so $\chi(a\cdot x)=-1$ for all $a\in \mathfrak{U}(R_{q})$.
Hence, by Proposition \ref{prop:44} we have
\[
w_{hom}(x)=\gamma\left[ 1-\frac{1}{\vert \mathfrak{U}(R_{q}\vert} \sum _{a\in  \mathfrak{U}(R_{q}}(-1)\right]=2\gamma.
\]
If $x \neq 0$ and $x \neq u_{1}u_{2}\cdots u_{q}$, then by Lemma \ref{lem:45}
we have $\sum_{a\in R_{q}} \chi(a \cdot x)=0$. Thus we obtain
\[w_{hom}(x)=\gamma\left[ 1-\frac{1}{\vert \mathfrak{U}(R_{q}\vert} 0\right]=\gamma.
\]
\qed

The homogeneous weight for a codeword $x=\left( x_{1},x_{2},\cdots, x_{n}\right) \in R_{q}^{n}$ is
defined as
\[
w_{hom}\left( x_{i}\right) =\left\{
\begin{array}{cl}
0 & \text{ \ \ if } x_{i}=0, \\
2^{q+1} & \text{ \ \ if } x_{i}=u_{1}u_{2} \cdots u_{q},\\
2^{q} & \text{ \ \ otherwise.}
\end{array}
\right.
\]

The corresponding Gray map is given by
\[
\begin{array}{ccccc}
\Psi _{hom} & : & R_{q} & \rightarrow & \mathbb{F}_{2}^{2^{^{q+1}}}
\end{array}
\]
where
\[
\begin{array}{ccc}
\Psi _{hom}\left( 0\right)& \text{=} & \text{}0000\cdots00 \\
\Psi _{hom}\left( 1\right)& \text{=} & \text{}0101\cdots01 \\
\vdots& \text{\vdots} & \text{}\vdots \\
\Psi _{hom}\left( \sum\limits
_{\substack{ A\subseteq \left\{ 1,2,\ldots,q\right\}  \\ c_{\varnothing }=0\vee A \neq
\varnothing}}c_{A}u_{A}\right)& \text{=} & \text{}1111 \cdots 11 \\
\end{array}
\]
Hence the following lemma holds.
\begin{lem}
If $C$ is a linear code over $R_q$ of length $n$, cardinality $2^k$
and minimum homogeneous weight $d_{hom}$,
then $\Psi_{hom}(C)$ is a binary linear code with parameters $[2^{2^{q+1}}n,k,d_{hom}]$.
\end{lem}

The following definition gives the Hamming, Lee and homogeneous weight distributions.
\begin{defi}
\cite{gupta}
For every $1 \leq i \leq n$, let $A_{Ham}(i)$, $A_{Lee}(i)$ and $A_{hom}(i)$
be the number of codewords of Hamming, Lee and homogeneous weight $i$ in $C$, respectively.
Then
\[
\begin{array}{l}
(A_{Ham}(0), A_{Ham}(1), \cdots, A_{Ham}(n)),\\
(A_{Lee}(0), A_{Lee}(1), \cdots, A_{Lee}(n)),\\
\mbox{ and }\\
(A_{hom}(0), A_{hom}(1), \cdots, A_{hom}(n)),
\end{array}
\]
are called the Hamming, Lee and homogeneous weight distributions of $C$, respectively.
\end{defi}

In \cite{Gulliver}, the {\it torsion} code of a code $C$ over $R_{q}$ was defined as
\begin{equation*}
Tor_{A}\left( C\right) =\left\lbrace v\in  \mathbb{F}_{2}^{n};\, u_{A}v \in C,A \subset \left\lbrace 1,
\cdots,2^{q}\right\rbrace \right\rbrace .\tag {1}
\end{equation*}
$Tor_{\varnothing}\left( C\right) =\left\lbrace v\in \mathbb{F}_{2}^{n};\, u_{\varnothing}v \in C,A =\varnothing \right\rbrace$
is called the residue code and is often denoted by
$Res(C)=\{u \in \mathbb{F}_{2}^{n};\exists v \in \mathbb{F}_{2}^{n};u+u_{A}v \in C\}$.
In general, we have the following tower of codes
\begin{equation*}
Tor_{\varnothing}\left( C\right)\subseteq Tor_{\{i\}\subset \left\lbrace 1,\cdots,2^{q}\right\rbrace}\left( C\right)\subseteq \cdots \subseteq Tor_{\left\lbrace 1,\cdots,2^{q}\right\rbrace}\left( C\right).
\tag{2}
\end{equation*}
Hence for a code $C$ over $R_{q}$
\begin{equation*}
\vert C \vert=\vert Tor_{\varnothing}\left( C\right)\vert \vert Tor_{\{i\}\subset \left\lbrace 1,\cdots,2^{q}\right\rbrace}\left( C\right) \vert \cdots \vert Tor_{\left\lbrace 1,\cdots,2^{q}\right\rbrace}\left( C\right) \vert.
\end{equation*}

Before presenting the simplex codes of types $\alpha$ and $\beta$, we define the $2$-dimension of a code $C$.
In \cite{Vazirani}, the authors presented the $p$-dimension for finitely generated modules over $\mathbb{Z}_{p^{s}}$.
Using this result, we define the $2$-dimension of a code $C$ over $R_{q}$ as follows.
A subset $S$ of $C$ is a $2$-basis for the linear code $C$ over $R_{q}$ if $S$
is $2$-linearly independent and $C$ is the 2-span of $S$.
The number of vectors in a $2$-basis for $C$ is called the $2$-dimension of $C$.
% and is denoted $2$-dim$(C)$.

\subsection{The Covering Radius}

The covering radius of a code is defined as the smallest integer $r$ such that all vectors in the space
are within distance $r$ of some codeword.
The covering radius of a code $C$ over $R_{q}$ is then
\[
r_{Lee}(C)= \max_{v\in R_{q}^{n}}\{d(v,C)\} \mbox{ and } r_{hom}(C)= \max_{v\in R_{q}^{n}}\{d(v,C)\},
\]
for the Lee and homogeneous weights, respectively.
It is easy to see that $r_{Lee}(C)$ and $r_{hom}(C)$ are the minimum values of $r_{Lee}$ and $r_{hom}$ such that
\[
R_{q}^{n } =\cup_{c \in C}
S_{r_{Lee}}(c) \mbox{ and }R_{q}^{n } =\cup_{c \in C}
S_{r_{hom}}(c),
\]
respectively, where
\[
S_{r_{Lee}}(u)=\left\lbrace  v\in R_{q}^{n } ; d(u,v)\leq r_{Lee} \right\rbrace \mbox{ and }
S_{r_{hom}}(u)=\left\lbrace  v\in R_{q}^{n } ; d(u,v)\leq r_{Lee} \right\rbrace.
\]

\begin{prop}\label{prop:33} \cite [Proposition 3.2]{Aoki}
Let $C$ be a code over $R_{q}^{n } $ and $\Psi_{Lee}(C)$ the Gray map image of $C$.
Then $r_{Lee}\left( C\right) =r_{Ham}( \Psi_{Lee}(C))$.
\end{prop}

\begin{prop}\label{prop:34}
If $C_{0}$ and $C_{1}$ are codes over $R_{q}^{n }$ generated by
matrices $G_{0}$ and $G_{1}$, respectively, and if $C$ is the code generated by
\[
G=\left[
\begin{array}{l|l}
0 & G_{1} \\ \hline
G_{0} & A
\end{array}
\right],
\]
then $r_{d}(C)\leq r_{d_{0}}(C_{0})+r_{d_{1}}(C_{1})$ and the covering radius of $C_c$
(the concatenation of $C_{0}$ and $C_{1}$)) satisfies the following inequality
$r_{d}(C_c)\geq
r_{d_{0}}(C_{0})+r_{d_{1}}(C_{1})$ for all distances $d$ over $R_{q}^{n }$.
\end{prop}

\pf
See \cite[Part D]{Cohen}.
\qed
\section{Simplex Codes of Type $\protect\alpha $}
Let $q$ and $k$ be positive integers with $q \geq 1$,
and let $G_{(q,k)}^{\alpha}$ be the matrix of size $k\times 2^{2^{q}\cdot k}$
defined inductively by
\begin{equation}
\label{equation:30}
G_{(q,k)}^{\alpha }=\left(
\begin{array}{l|l|l}
G_{(q,k-1)}^{\alpha } & \ldots& G_{(q,k-1)}^{\alpha} \\
\hline
0_{2^{2^{q} \cdot \left( k-1\right) }} & \ldots & \left( 1+\sum\limits_{\substack{ A\subseteq \left\{ 1,2,\ldots,q\right\}  \\
c_{\varnothing}=0\vee A\neq \varnothing }}c_{A}u_{A}\right) \times 1_{_{2^{2^q \cdot \left( k-1\right) }}}
\end{array}
\right),
\tag{3}
\end{equation}
for $k\geq 2$, where
\[
G_{(q,1)}^{\alpha}=\left(0\ 1\ u_{1} \ \cdots \ \left( 1+\sum\limits_{\substack{ A\subseteq \left\{ 1,2,\ldots,q\right\}  \\
c_{\varnothing}=0\vee A\neq \varnothing }}c_{A}u_{A}\right)\right),
\]
is a matrix with one row and $2^{2^q}$ columns containing all the elements of $R_{q}$.
The columns of $G_{(q,k)}^{\alpha}$ consist of all distinct $k$-tuples over $R_{q}$.
The code $S_{(q,k)}^{\alpha}$ generated by $G_{(q,k)}^{\alpha}$ is called the simplex code of type $\alpha$ over $R_q$.
This code has length $2^{2^{q}k}$ and $2$-dimension $2^{q}k$.

\begin{rem}\label{rem:1}
If $A_{k-1}$ denotes the $2^{2^q \cdot (k-1)}\times 2^{2^q \cdot (k-1)}$
matrix consisting of all codewords in $S_{(q,k-1)}^{\alpha }$,
and $J$ is the matrix with all elements equal to $1$,
then $S_{(q,k)}^{\alpha}$ is generated by the
$2^{2^q \cdot k} \times 2^{2^q \cdot k}$ matrix
\begin{equation}
\left[
\begin{array}{cccc}
A_{k-1} & A_{k-1} & \cdots & A_{k-1} \\
A_{k-1} & J+A_{k-1} & \cdots & \left( 1+\sum\limits_{\substack{ A\subseteq
\left\{ 1,2,\ldots ,q\right\}  \\ c_{\varnothing }=0\vee A\neq \varnothing }}c_{A}u_{A}\right)
J+A_{k-1} \\
\vdots & \vdots & \ddots & \vdots \\
A_{k-1} & \left( 1+\sum\limits_{\substack{ A\subseteq \left\{
1,2,\ldots,q\right\}  \\ c_{\varnothing }=0\vee A\neq \varnothing }}c_{A}u_{A}\right)
J+A_{k-1} & \cdots & J+A_{k-1}%
\end{array}
\right].  \tag{4}
\end{equation}
\end{rem}

\begin{rem}
If $l_{1},l_{2},\ldots,l_{k}$ are the rows of $G_{(q,k)}^{\alpha }$, then
\begin{enumerate}
\item
$w_{Ham}\left(l_{i}\right)
=3 \cdot 5\cdot 17 \cdot 257 \cdot \ldots \cdot 2^{2^q \cdot k - 2^q}$,
$w_{Ham}\left( u_{1}l_{i}\right) =w_{Ham}\left( u_{2}l_{i}\right) = \ldots = w_{Ham}\left( u_{q}l_{i}\right)
=3\cdot 5\cdot 17\cdot 257 \cdot \ldots \cdot 2^{2^q \cdot k-2^{q-1}}$,
$w_{Ham}\left(u_{1}u_{2}\ldots u_{q}l_{i}\right) =2^{2^q \cdot k-1}$.

\item
$w_{Lee}\left( l_{i}\right) =w_{Lee}\left( u_{1}l_{i}\right) =w_{Lee}\left(
u_{2}l_{i}\right) = \ldots=w_{Lee}\left( u_{1}u_{2}\ldots u_{q}l_{i}\right)
=2^{2^q \cdot k+(q-1)}$.

\item$w_{hom}\left( l_{i}\right) =w_{hom}\left( u_{1}l_{i}\right) =w_{hom}\left(
u_{2}l_{i}\right) = \ldots=w_{hom}\left( u_{1}u_{2}\ldots u_{q}l_{i}\right)
=2^{2^{q}k}$.
\end{enumerate}
\end{rem}

In the matrix $G_{(q,k)}^{\alpha }$, it is clear that each element of $R_{q}$ appears $2^{2^q\cdot (k-1)}$ times in
every row.
Thus we have the following lemma.

\begin{lem}
\label{lem:3}
Let $c\in S_{(q,k)}^{\alpha }$ be nonzero.
If one coordinate of $c$ is a unit then every element of $R_{q}$ occurs $2^{2^q\cdot (k-1)}$ times
as a coordinate of $c$.
\end{lem}

\pf
%\begin{proof}
From Remark \ref{rem:1}, any $x\in S_{(q,k-1)}^{\alpha }$ gives the following
codewords of $S_{(q,k)}^{\alpha}$
\[
\begin{array}{ccl}
c_{1}&=&\left(
x \vert x \vert x \vert \cdots \vert x
\right)\\
c_{2}&=&\left(
x \vert 1+x \vert u_{1}+x \vert \cdots \vert \left( 1+\sum\limits_{\substack{ A\subseteq \left\{ 1,2,\ldots ,q\right\} \\
c_{\varnothing }=0\vee A\neq \varnothing}} c_{A}u_{A}\right) +x
\right)\\
\vdots&& \\
c_{2^{2^{^{q}}}}&=&
\left(
x \vert \left( 1+\sum\limits_{\substack{ A\subseteq \left\{1,2,\cdots,q\right\} \\
c_{\varnothing }=0 \vee A \neq \varnothing }}c_{A}u_{A}\right) + x \vert \cdots \vert x \right).
\end{array}
\]
The result then follows by induction on $k$ and Remark \ref{rem:1}.
\qed

To obtain the torsion codes over $R_q$, it is necessary to introduce the binary simplex codes of type $\alpha$ and $\beta$.

The binary simplex code of type $\alpha$, denoted by $S_k$, has parameters $[2^{k};k;d_{Ham}=2^{k-1}]$ and generator matrix
\begin{equation*}
\label{equation*:11}
G_{k}=\left(
\begin{array}{l|l}
00\cdots 0 & 11\cdots 1 \\
\hline
G_{k-1} & G_{k-1}
\end{array}
\right), \tag{5}
\end{equation*}
for $k \geqslant 2$, where  $G_{1}=\left( 0 | 1 \right)$.

The binary simplex code of type $\beta$, denoted by $\widehat{S}_{k}$, has parameters $[2^{k}-1;k;d_{Ham}=2^{k-1}]$ and generator matrix
\begin{equation*}
\label{equation*:33}
\widehat G_{k}=\left(
\begin{array}{l|l}
11\cdots 1 & 00\cdots 0 \\
\hline
G_{k-1} & \widehat G_{k-1}
\end{array}
\right), \tag{6}
\end{equation*}
for $k \geqslant 3$, where
 \begin{equation*}
\widehat G_{2}=\left(
\begin{array}{l|l}
11 & 0 \\
\hline
01 & 1
\end{array}
\right).
\end{equation*}
\begin{lem}\label{lem:hh}
The torsion code of $S_{(q,k)}^{\alpha}$ is the concatenation of $2^{(2^q-1)k}$ ${S}_{k}$ codes.
\end{lem}

\pf
The torsion code of $S_{(q,k)}^{\alpha }$ is the set of codewords obtained by
replacing
$u_{1}u_{2}\cdots u_{q}$ with
$1$ in all $u_{1}u_{2}\ldots u_{q}$-linear combinations of the rows of $u_{1}\cdots u_{q}G_{(q,k)}^{\alpha}$
(where $G_{(q,k)}^{\alpha}$ is the generator matrix of $S_{(q,k)}^{\alpha}$ defined in (\ref{equation:30})).
The proof is by induction on $k$.
For $k=2$, the result is true. If $u_{1}u_{2}\cdots u_{q}G_{(q,k-1)}^{\alpha}$ is the matrix obtained by
the concatenation of $2^{(2^q-1)(k-1)}$ copies of the matrix $u_{1}u_{2}\cdots u_{q}G_{k-1}$,
then $u_{1}u_{2}\cdots u_{q}G_{(q,k)}^{\alpha}$ takes the form
\begin{equation*}
\left(
\begin{array}{l|l|l}
 u_{1}u_{2}\cdots u_{q}G_{k-1} \cdots
  u_{1u_{2}}\cdots u_{q}G_{k-1} & \cdots &
u_{1}u_{2}\cdots u_{q}G_{k-1} \cdots
u_{1}u_{2}\cdots u_{q}G_{k-1} \\ \hline
0_{_{_{2^{2^{^{q}}\cdot \left( k-1\right) }}}} & \cdots & (u_{1}u_{2}\ldots
u_{q})\times 1_{_{_{2^{2^{^{q}}\cdot \left( k-1\right) }}}}
\end{array}
\right).
\tag{7}
\end{equation*}
Grouping the columns based on (\ref{equation*:11}), we obtain the result.
\qed

For $q\geq 2$, we define the following linear homomorphism
\begin{equation*}
\begin{array}{ccccc}
\Gamma _{q} & : & R_{q} & \rightarrow & R_{q-1} \\
&  & 1+\sum\limits_{\substack{ A\subseteq \left\{ 1,2, \cdots ,q\right\}  \\
c_{\varnothing }=0\vee A\neq \varnothing}}c_{A}u_{A} & \longmapsto & \Gamma _{q}\left(
1+\sum\limits_{\substack{ A\subseteq \left\{ 1,2,\cdots ,q\right\}  \\ c_{\varnothing
}=0\vee A\neq \varnothing }}c_{A}u_{A}\right),
\end{array}
\end{equation*}
where
\begin{equation*}
\Gamma _{q}\left( 1+\sum\limits_{\substack{ A\subseteq \left\{
1,2,\cdots,q\right\}  \\ c_{\varnothing }=0\vee A\neq \varnothing}}c_{A}u_{A}\right)
=1+\sum\limits_{\substack{ A\subseteq \left\{ 1,2,\cdots,q-1\right\}.  \\
c_{\varnothing }=0\vee A\neq \varnothing}}c_{A}u_{A}.
\end{equation*}
We have
\begin{equation*}
Im(\Gamma_{q})=R_{q-1},
\end{equation*}
and for $n$ a positive integer this homomorphism can be extended to $R_{q}^{n}$
\begin{equation*}
\begin{array}{ccccc}
\Gamma _{q} & : & R_{q}^{n} & \longrightarrow & R_{q-1}^{n}.
\end{array}
\end{equation*}

\begin{thm}\label{thm:11}
Let $S_{(q,k)}^{\alpha}$ be the simplex code of type $\alpha$ over $R_{q}$.
Then $\Gamma _{q}(S_{(q,k)}^{\alpha })$ is the concatenation of $2^{2^{q-1} k}$
simplex codes of type $\alpha$ over $R_{q-1}$.
\end{thm}
\pf
If $G_{(q,k)}^{\alpha}$ is a generator matrix of the simplex code $S_{(q,k)}^{\alpha}$ of type $\alpha$
over $R_{q}$, then $\Gamma_{q}(G_{(q,k)}^{\alpha})$ has the form
\begin{equation*}
\Gamma _{q}(G_{(q,k)}^{\alpha })=\left( \overset{2^{2^{q-1}k}}{\overbrace{
\begin{array}{l|l|l|l}
G_{(q-1,k)}^{\alpha } & G_{(q-1,k)}^{\alpha } & \cdots  &
G_{(q-1,k)}^{\alpha }
\end{array}
}}\right),
\end{equation*}
where
\begin{equation*}
G_{(q-1,k)}^{\alpha }=\left(
\begin{array}{l|l|l|l}
G_{(q-1,k-1)}^{\alpha } & G_{(q-1,k-1)}^{\alpha } & \cdots &G_{(q-1,k-1)}^{\alpha } \\
\hline
0_{_{2^{2^{q-1}\left( k-1\right) }}} & 1_{2^{2^{q-1}\left( k-1\right) }} & \cdots
& \left( 1+\sum\limits_{\substack{ A\subseteq \left\{ 1,2,\ldots,q\right\}
\\ c_{\varnothing }=0\vee A\neq \varnothing }}c_{A}u_{A}\right) \times 1_{_{2^{2^{q-1}\left(k-1\right) }}}
\end{array}
\right),
\end{equation*}
is a generator matrix of the simplex code of type $\alpha$ over $R_{q-1}$.
\qed

\begin{thm}
\label{thm:3}
If $S_{(q,k)}^{\alpha}$ is a simplex code of type $\alpha $ over $R_{q}$, then
\[
\Gamma _{q}\left(\Gamma _{q-1}\cdots \left( \Gamma _{2}\left( S_{(2,k)}^{\alpha }\right) \right)
\right)
=\left( \overset{2^{2^{q\left( \frac{q-1}{q}\right) }k}}{\overbrace{S_{(1,k)}^{\alpha }S_{(1,k)}^{\alpha }\cdots S_{(1,k)}^{\alpha }}}\right),
\]
is the concatenation of $2^{2^{q\left( \frac{q-1}{q}\right) }k}$ $S_{(1,k)}^{\alpha}$ codes
where $S_{(1,k)}^{\alpha }$ is the simplex code of type $\alpha $ over $R_{1}$.
\end{thm}
\pf
The proof is by induction on $q$ and Theorem~\ref{thm:3}.
For $q=2$, if $ G_{(2,k)}^{\alpha}$ is a generator matrix of the simplex code over $R_{2}$, then
\[
\Gamma_{2}\left( G_{(2,k)}^{\alpha}\right) =\left( \overset{2^{2k}}{\overbrace{
\begin{array}{l|l|l|l}
G_{(1,k)}^{\alpha} & G_{(1,k)}^{\alpha} & \cdots & G_{(1,k)}^{\alpha}
\end{array}
}}\right),
\]
where $G_{(1,k)}^{\alpha}$ is a generator matrix of the simplex code over $R_{1}$.
If
\[
\Gamma _{q-1}\left(\Gamma _{q-2}\cdots \left( \Gamma _{2}\left(  G_{(2,k)}^{\alpha}\right) \right)
\right)=\left( 2^{2^{q-1}k}\cdots\left( 2^{2^{2}k}G_{(1,k)}^{\alpha}\right)\right),
\]
is the generator matrix obtained by the concatenation of $2^{2^{(q-2)\left( \frac{q+1}{2}\right) }k}$ simplex codes of type $\alpha$ over $R_{1}$. Then
\[
\Gamma _{q}\left(\Gamma _{q-1}\cdots \left( \Gamma _{2}\left(  G_{(2,k)}^{\alpha}\right) \right)\right)=\left( 2^{2^{q}k}\cdots\left( 2^{2^{2}k} G_{(1,k)}^{\alpha}\right)\right)
=\left( \overset{2^{2^{q\left( \frac{q-1}{2}\right) }k}}{\overbrace{
\begin{array}{l|l|l|l}
 G_{(1,k)}^{\alpha} & G_{(1,k)}^{\alpha} & \cdots  & G_{(1,k)}^{\alpha}
\end{array}
}}\right).
\]
\qed

Let $S_{0}=\left\lbrace 0\right\rbrace $,
$S_{1}=\left\lbrace 0,u_{1}u_{2}\cdots u_{q}\right\rbrace $,$\cdots$,
$S_{q-1}=\left\lbrace 0,u_{1},u_{2},\cdots,u_{1}u_{2}\cdots u_{q}\right\rbrace$,
and $S_{q}=R_{q}$.
Note that $S_{q-1}$ is the set of all zero divisors of $R_{q}$.
A codeword $c=(c_{1},c_{2},\cdots,c_{n})\in S_{(q,k)}^{\alpha}$ is said to
be of {\it type $m$}, $0\leq m \leq q$, if all of its components belong to the set $S_{m}$.
From $G_{(q,k)}^{\alpha}$, we have that each element of $R_{q}$ occurs equally often in every row of $G_{(q,k)}^{\alpha}$.

To determine the Hamming, Lee and homogeneous weight distributions of $S_{(q,k)}^{\alpha}$,
the number of codewords of type $m$ in $S_{(q,k)}^{\alpha }$, $0\leq m\leq q$, must be determined.
For this, we define the matrix $D_{i}$ as
\begin{equation*}
D_{0}=\left(
\begin{array}{c}
u_{1}u_{2}\ldots u_{q}l_{1} \\
u_{1}u_{2}\ldots u_{q}l_{2} \\
\vdots \\
u_{1}u_{2}\ldots u_{q}l_{k}%
\end{array}
\right), \;
D_{1}=\left(
\begin{array}{c}
u_{1}\ldots u_{q-1}l_{1} \\
u_{1}\ldots u_{q}l_{1} \\
u_{1}\ldots u_{q-1}l_{2} \\
u_{1}\ldots u_{q}l_{2} \\
\vdots \\
u_{1}\ldots u_{q-1}l_{k} \\
u_{1}\ldots u_{q}l_{k}
\end{array}
\right), \;
\cdots, \;
D_{q}=\left(
\begin{array}{c}
l_{1} \\
u_{1}l_{1} \\
\vdots \\
u_{1}\ldots u_{q}R_{1} \\
l_{2} \\
u_{1}l_{2} \\
\vdots \\
u_{1}\ldots u_{q}l_{2} \\
\vdots \\
l_{k} \\
u_{1}l_{k} \\
\vdots \\
u_{1}\ldots u_{q}l_{k}
\end{array}
\right),
\end{equation*}
where $l_{i}$ is the $i^{th}$ row of $G_{(q,k)}^{\alpha}$. Let $C^{(m)}$ be the subcode of $C$ generated by the rows of $D_{m}$.
We then have that
\begin{equation*}
C^{(0)}\subset C^{(2)}\subset \cdots \subset C^{(q)}.
\end{equation*}
Note that $C^{(m)}$ has $2^{mk}$ codewords and the matrix $D_{q}$ generates $S_{(q,k)}^{\alpha}$.
For $0\leq m\leq q$, the codewords of type $m$ occur
$2^{mk}-2^{(m-1)k}$ times in $S_{(q,k)}^{\alpha}$.
This proves the following lemma.

\begin{lem}\label{lem:2}
For $0\leq m\leq q$, the number of codewords of type $m$ in $S_{(q,k)}^{\alpha}$ is $2^{(m-1)k}(2^{k}-1)$.
\end{lem}

\begin{thm}\label{thm:7}
The Hamming, Lee and homogeneous weight distributions of $S_{(q,k)}^{\alpha }$ are
\begin{enumerate}
\item[(i)]
$A_{Ham}(0)=1$,
$A_{Ham}(2^{(^{2^{q}k-m)}})(2^{m}-1))=2^{(m-1)k}(2^{m}-1)$, for $0\leq m\leq q$.

\item[(ii)]
$A_{Lee}(0)=1,
A_{Lee}(2^{2^q k+(q-1)})=2^{2^qk}-1$.

\item[(iii)]
$A_{hom}(0)=1,
A_{hom}(2^{2^{q}k})=2^{2^qk}-1$.

\end{enumerate}
\end{thm}

\pf
Let $c \in S_{(q,k)}^{\alpha }$ be a codeword of type $m$ $\neq 0$.
Then by Lemma \ref{lem:2}
\[
A_{Ham}(2^{2^q-m}(2^{m}-1))=2^{(m-1)k}(2^{m}-1),
\]
for $m=0$, and $A_{Ham}(0)=1$.
Further, by Lemma \ref{lem:3},
$A_{Lee}(c)=2^{2^q \cdot k}-1$ which is independent of $m$,
so all codewords of type $m \neq 0$ have the same Lee and homogeneous weights.
\qed

\subsection{Binary Gray Images of Simplex Codes of Type $\alpha$}

The binary images of the simplex code $S_{(q,k)}^{\alpha}$ over $R_{q}$ are
given in the following two theorems.
\begin{thm}
\label{thm:4}
Let $S_{(q,k)}^{\alpha }$ be the simplex code over $R_{q}$ of length $2^{2^{q}k}$, 2-dimension $2^{q}k$,
and minimum Lee weight $d_{Lee}$.
Then $\Psi _{Lee}(S_{(q,k)}^{\alpha })$ is the concatenation of $2^{(2^{q}-1)k+q}$ binary simplex codes with
parameters
$[2^{2^{q}k+q};k;d_{Ham}=2^{2^{q}k+q-1}]$.
\end{thm}
\pf
Let $G_{(q,k)}^{\alpha}$ be a generator matrix of the simplex code
$S_{(q,k)}^{\alpha}$ over $R_q$.
Then $\Psi _{Lee}(G_{(q,k)}^{\alpha })$ has the form
\begin{equation*}
\Psi _{Lee}(G_{(q,k)}^{\alpha })=\left( \overset{2^{(2^{q}-1)k+q}}{\overbrace{
\begin{array}{l|l|l|l}
G_{k} & G_{k} & \cdots & G_{k}
\end{array}
}}\right),
\end{equation*}
where $G_{k}$ is a generator matrix of the binary simplex code $S_{k}$.
The result then follows by induction on $k$.
\qed

\begin{thm}
\label{thm:15}
Let $S_{(q,k)}^{\alpha}$ be the simplex code over $R_{q}$ of length $2^{2^{q}k}$, 2-dimension $2^{q}k$,
and minimum homogeneous weight $d_{hom}$.
Then $\Psi_{hom}(S_{(q,k)}^{\alpha})$ is the concatenation of
$2^{(2^{q}-1)k+q+1}$ binary simplex codes with
parameters $[2^{2^{q}k+q+1};k;d_{Ham}=2^{2^{q}k+q}]$.
\end{thm}
\pf
The proof is similar to that of Theorem \ref{thm:4}.
\qed

\section{Simplex Codes of Type $\beta$}

Let $G_{(q,k)}^{\beta}$ be the matrix of size $k\times 2^{\left( 2^{2^{q}}-1\right)
(k-1)}(2^{k}-1)$ defined by
\begin{equation*}
G_{(q,k)}^{\beta }=\left(
\begin{array}{l|l|l|l}
1_{2^{2^{^{q}}\cdot \left( k-1\right) }} & 0_{2^{2^{^{q}}\left(k-2\right) }\left( 2^{k-1}-1\right) } & \ldots &
\left( \sum_{\substack{ A\subseteq \left\{ 1,2,\ldots,q\right\}  \\
c_{\varnothing }=0\vee A\neq \varnothing }} c_{A}u_{A}\right)_{2^{2^{^{q}}\left( k-2\right) }\left( 2^{k-1}-1\right)} \\
\hline
G_{(q,k-1)}^{\alpha} & G_{(q,k-1)}^{\beta} & \ldots & G_{(q,k-1)}^{\beta}
\end{array}
\right),
\end{equation*}
for $k > 2$ and
\begin{equation*}
G_{(q,2)}^{\beta }=\left(
\begin{array}{l|l|l|l}
1_{_{2^{2^{^{q}}\cdot \left( k-1\right) }}} & 0 & \ldots & \sum
_{\substack{ A\subseteq \left\{ 1,2,\ldots,q\right\} \\
c_{\varnothing }=0\vee A\neq \varnothing}}c_{A}u_{A} \\
\hline
0\ 1\ \ldots\left( 1+\sum_{\substack{ A\subseteq \left\{ 1,2,\ldots,q\right\} \\
c_{\varnothing }=0\vee A\neq \varnothing}}c_{A}u_{A}\right) & 1 & \ldots & 1
\end{array}
\right),
\end{equation*}
where $G_{(q,k-1)}^{\alpha}$ is a generator matrix of $S_{(q,k-1)}^{\alpha}$.

\begin{rem}\label{rem:3}
Let $A_{k-1}$ $(B_{k-1})$ denote the array of codewords in $S_{(q,k-1)}^{\alpha}$
$(S_{(q,k-1)}^{\beta })$, and $J$ the matrix of all 1's.
Then the array of codewords of $S_{(q,k)}^{\beta }$ is given by the following matrix
\end{rem}
\begin{equation*}
\left[
\begin{array}{cccc}
A_{k-1} & B_{k-1} & \cdots & B_{k-1} \\
J+A_{k-1} & B_{k-1} & \cdots & \left( \sum\limits_{\substack{ A\subseteq
\left\{ 1,2,\ldots,q\right\}  \\ c_{\varnothing }=0\vee A\neq \varnothing }}c_{A}u_{A}\right)
J+B_{k-1} \\
\vdots & \vdots & \ddots & \vdots \\
\left( 1+\sum\limits_{\substack{ A\subseteq \left\{ 1,2,\ldots,q\right\}  \\
c_{\varnothing }=0\vee A\neq \varnothing }}c_{A}u_{A}\right) J+A_{k-1} & B_{k-1} & \cdots
& \left( \sum\limits_{\substack{ A\subseteq \left\{ 1,2,\ldots,q\right\}  \\
c_{\varnothing }=0\vee A\neq \varnothing}}c_{A}u_{A}\right) J+B_{k-1}
\end{array}
\right].
\end{equation*}

Let $\mathfrak{U(R_{q})}$ and $\mathfrak{D(R_{q})}$ denote the set of units and the set of zero divisors of $R_{q}$, respectively.
The following proposition provides the weight distributions of $S_{(q,k)}^{\beta}$.
\begin{prop}
For $1\leq j\leq k$, let $l_{j}$ be the $j${th} row of $G_{(q,k)}^{\beta }$.
Then we have
\begin{enumerate}
\item[(i)] $\sum\limits_{i\in \mathfrak{U(R_{q})}}w_{i}=2^{2^{^{q}}\cdot (k-1)}$ and each zero
divisor in $R_{q}$ appears $2^{(2^q-1)\cdot (k-2)}(2^{k-1}-1)$ times in $l_{j}$.

\item[(ii)] $w_{Ham}(l_{j})=2^{(2^q-1)(k-1)-2^q}(3 \cdot 5 \cdot 17 \cdot 257 \cdots (2^{k}-1)+1)$.

\item $w_{Lee}(l_{1})=2^{2^q(k-1)}+2^{2^q\cdot k-(2^q-1)}-2^{4k-(2^q-2)}$.

\item[(iii)] $w_{hom}(l_{j})=2^{(^{2^{q}-1)k-1}}(2^{k}-1).$
\end{enumerate}
\end{prop}

\pf
The proof follows from the definition of $l_{j}$.
\qed

The following proposition gives the structure of the codewords of $S_{(q,k)}^{\beta}$.
\begin{prop}
Consider a codeword $c\in S_{(q,k)}^{\beta}$.
If one coordinate of $c$ is a unit then $\sum_{i\in \mathfrak{U(R_{q})}}w_{i}=2^{2^{^{q}}\cdot (k-1)}$,
and each zero divisor in $R_{q}$ appears $2^{(2^{^{q}}-1)\cdot (k-2)}(2^{k-1}-1)$ times in $c$.
\end{prop}

\pf
By Remark \ref{rem:3}, there exists $x_{1}\in S_{(q,k-1)}^{\alpha }$ and $x_{2}\in
S_{(q,k-1)}^{\beta }$ such that $c$ takes one of following $2^{2^q}$ forms
\[
\begin{array}{ccl}
c_{1}&=&\left(
x_{1} \vert x_{2} \vert x_{2} \vert\cdots \vert x_{2} \right)\\
c_{2}&=&\left( 1+x_{1} \vert x_{2} \vert u_{1}+x_{2} \vert \cdots \vert \left( \sum\limits_{\substack{ A\subseteq \left\{ 1,2,\ldots,q\right\} \\
c_{\varnothing }=0\vee A\neq \varnothing }}c_{A}u_{A}\right) +x_{2}
\right)\\
\vdots &&\\
c_{2^{2^{^{q}}}}&=&\left( \left( 1+\sum\limits_{\substack{ A\subseteq \left\{ 1,2,\ldots,q\right\}  \\
c_{\varnothing } = 0\vee A\neq \varnothing }}c_{A}u_{A}\right) +x_{1} \vert \cdots \vert
\left(\sum\limits_{\substack{ A\subseteq \left\{ 1,2,\ldots,q\right\} \\
c_{\varnothing}=0\vee A\neq \varnothing }}c_{A}u_{A}\right) +x_{2}
\right).
\end{array}
\]
The result then follows by induction on $k$.
\qed

\begin{lem}
The torsion code of $S_{(q,k)}^{\beta }$ is the concatenation of $2^{(2^q-1)\cdot (k-2)}$
binary simplex codes of type $\beta$ denoted by $\widehat{S}_{k}$.
\end{lem}
\pf
The proof is similar to that of Lemma \ref{lem:hh}.
\qed
\begin{thm}
The Hamming and homogeneous weight distribution of $S_{(q,k)}^{\beta }$ are
\begin{enumerate}
\item[(i)] $A_{Ham}(0)=1$,
$A_{Ham}(2^{(2^q-1)(k-1)}[(2^{k-m}(2^{m}-1)+(2^{1-m}-1)])=2^{(m-1)k}(2^{m}-1)$, $0\leq m\leq q$.

\item[(ii)] $A_{hom}(0)=1$, $A_{hom}(2^{(2^{q}-1)k-1}(2^{k}-1)=2^{k}(2^{(2^{q}-1)k}-1)$.
\end{enumerate}
\end{thm}
\pf
The proof is similar to that of Theorem \ref{thm:7}.
\qed

\begin{thm}
\label{thm:5}
Let $S_{(q,k)}^{\beta }$ be the simplex code of type $\beta$ over $R_{q}$.
Then $\Gamma_{q}(S_{(q,k)}^{\beta})$ is the concatenation of $2^{2^{q-1}k}$ simplex codes of type $\beta$
over $R_{q-1}$.
\end{thm}

\pf If $G_{(q,k)}^{\beta}$ is a generator matrix of the simplex code of type $\beta$ over $R_{q}$,
then $\Gamma_{q}(G_{(q,k)}^{\beta})$ has the form
\begin{equation*}
\Gamma _{q}(G_{(q,k)}^{\beta })=\left( \overset{2^{2k }
}{\overbrace{
\begin{array}{l|l|l|l}
G_{(q-1,k)}^{\beta } & G_{(q-1,k)}^{\beta } & \cdots  & G_{(q-1,k)}^{\beta }
\end{array}
}}\right) ,
\end{equation*}
where $G_{(q-1,k)}^{\beta} $ is a generator matrix of the simplex code $S_{(q,k-1)}^{\beta }$ of type $\beta$ over $R_{q-1}$.
\qed

\begin{thm}
\label{thm:12}
If $S_{(q,k)}^{\beta}$ is the simplex code of type $\beta$ over $R_{q}$,
then
\[
\Gamma _{q}\left( \Gamma _{q-1}\cdots \left( \Gamma _{2}\left( S_{(2,k)}^{\beta}\right) \right) \right)
=\left( \overset{2^{2^{q\left( \frac{q-1}{2}\right)} k }}{\overbrace{S_{(1,k)}^{\beta}S_{(1,k)}^{\beta}\cdots
S_{(1,k)}^{\beta}}}\right),
\]
is the concatenation of $2^{2^{q\left( \frac{q-1}{2}\right) }k }$ simplex codes of type $\beta$ over $R_{1}$,
denoted by $S_{(1,k)}^{\beta}$.
\end{thm}

\pf
The proof is by induction on $q$ and Theorem \ref{thm:5}.
For $q=2$, $G_{(2,k)}^{\beta}$ is a generator matrix for the simplex codes of type $\beta$ over $R_{2}$.
Then
\[
\Gamma _{2}\left(
G_{(2,k)}^{\beta}\right) =\left( \overset{2^{2k} }{\overbrace{
\begin{array}{l|l|l|l}
G_{(1,k)}^{\beta} & G_{(1,k)}^{\beta} & \cdots & G_{(1,k)}^{\beta}
\end{array}
}}\right),
\]
where $G_{(1,k)}^{\beta}$ is a generator matrix for the simplex code of type $\beta$ over $R_{1}$.
Then
\[
\Gamma_{q-1}\left( \Gamma_{q-2}\cdots \left( \Gamma_{2}\left( G_{(2,k)}^{\beta}\right) \right) \right)
=\left( 2^{2^{q-1}k}\cdots \left( 2^{2^{2}k}G_{(1,k)}^{\beta}\right) \right),
\]
is the generator matrix obtained by the concatenation of  $2^{2^{q\left( \frac{q-1}{2}\right)}k}$  $S_{(1,k)}^{\beta}$ codes,
where $S_{(1,k)}^{\beta}$ is the simplex code of type $\beta$ over $R_{1}$.
Then
\[
\Gamma_{q}\left( \Gamma_{q-1}\cdots \left( \Gamma_{2}\left(
S_{(2,k)}^{\beta }\right) \right) \right) =\left( \overset{2^{2^{(q-2)\left( \frac{q+1}{2}\right)}k }}{\overbrace{
\begin{array}{l|l|l|l}
S_{(1,k)}^{\beta} & S_{(1,k)}^{\beta} & \cdots & S_{(1,k)}^{\beta}
\end{array}
}}\right).
\]
\qed

\subsection{Binary Gray Images of the Simplex Codes of Type $\beta$}

The binary images of the simplex codes of type $\beta$ over $R_{q}$ are given in the following theorems.

\begin{thm}
\label{thm:6}
Let $S_{(q,k)}^{\beta}$ be the simplex code over $R_{q}$ of length $
2^{( 2^{2^{q}}-1)(k-1)}(2^{k}-1)$, 2-dimension $2^{q}k$ and minimum Lee weight
$d_{Lee}$.
Then $\Psi_{Lee}(S_{(q,k)}^{\beta })$ is the concatenation of
$2^{\left( 2^{2^{q}}-1\right)(k-1)+q}$ simplex codes with parameters
$[2^{( 2^{2^{q}}-1)(k-1)+q}(2^{k}-1);k;d_{Ham}=2^{(2^{q-1}-2)k+q}]$.
\end{thm}
\pf
If $G_{(q,k)}^{\beta}$ is a generator matrix of the simplex code $S_{(q,k)}^{\beta}$ over $R_q$,
then $\Psi _{Lee}(G_{(q,k)}^{\alpha })$ has the following form
\begin{equation*}
\Psi _{Lee}(G_{(q,k)}^{\alpha })=\left( \overset{2^{\left( 2^{2^{q}}-1\right)(k-1)+q}}{
\overbrace{
\begin{array}{l|l|l|l}
G_{k} & G_{k} & \cdots & G_{k}
\end{array}
}}\right),
\end{equation*}
where $G_{k}$ is a generator matrix of the binary simplex code $S_{k}$.
The result then follows by induction on $k$.
\qed

\begin{thm}
\label{thm:16}
Let $S_{(q,k)}^{\beta }$ be the simplex code over $R_{q}$ of length $
2^{( 2^{2^{q}}-1)(k-1)}(2^{k}-1)$, 2-dimension $2^{q}k$ and minimum homogeneous weight $d_{hom}$.
Then $\Psi_{\hom}(S_{(q,k)}^{\beta})$ is the concatenation of $2^{\left( 2^{2^{q}}-1\right)(k-1)+(q+1)}$ binary simplex codes with parameters
$[2^{( 2^{2^{q}}-1)(k-1)+(q+1)}(2^{k}-1);k;d_{Ham}=2^{\left( 2^{2^{q}}-2\right)(k-1)+(q+1)}]$.
\end{thm}

\pf
The proof is similar to that of Theorem \ref{thm:6}.
\qed

\section{MacDonald Codes of Types $\alpha$ and $\beta$ over ${R}_{q}$}

In \cite{Patel}, the MacDonald code $\mathcal{M}_{k,u}(q)$ over the finite field
$\mathbb{F}_{q}$ was defined as the unique $\left[ \frac{q^{k}-q^{u}}{q-1},k,q^{k-1}-q^{u-1}
\right] $ code in which every nonzero codeword has weight either $q^{k-1}$ or $q^{k-1}-q^{u-1}$.

Let $G_{(q,k)}^{\alpha }$ and $G_{(q,k)}^{\beta}$
be the generator matrices of the simplex codes of types $\alpha$ and $\beta$ over $R_q$, respectively.
For $1 \leq u \leq k-1$,
we define $G_{(q,k,u)}^{\alpha}$ (resp. $G_{(q,k,u)}^{\beta}$),
as the generator matrix of the MacDonald code $\mathcal{M}_{(q,k,u)}^{\alpha}$ (resp. $\mathcal{M}_{(q,k,u)}^{\beta}$),
obtained from $G_{(q,k)}^{\alpha }$ ( resp. $G_{(q,k)}^{\beta}$), by deleting
the columns corresponding to the columns of $G_{(q,u)}^{\alpha }$ and $0_{2^{2^{q}u}\times
\left( k-u\right) }$ (resp. $G_{(q,u)}^{\beta}$ and $0_{2^{(2^{q}-1)(u-1)}(2^{u}-1) \times \left(
k-u\right)})$, given by
\begin{equation}\label{equation:44}
G_{(q,k,u)}^{\alpha }=\left(
\begin{array}{ccc}
G_{(q,k)}^{\alpha} & \left\backslash {}\right. & \dfrac{0_{2^{2^{q}u}\times
\left( k-u\right) }}{G_{(q,u)}^{\alpha }}
\end{array}
\right),
\tag{8}
\end{equation}
and
\begin{equation}
G_{(q,k,u)}^{\beta }=\left(
\begin{array}{ccc}
G_{(q,k)}^{\beta } & \left\backslash {}\right. & \dfrac{0_{2^{(2^{q}-1)(u-1)}(2^{u}-1) \times \left(
k-u\right)} }{G_{(q,u)}^{\beta }}
\end{array}
\right).
\tag{9}
\end{equation}
The code $\mathcal{M}_{(q,k,u)}^{\alpha}$ (resp. $\mathcal{M}_{(q,k,u)}^{\beta}$),
generated by $G_{(q,k,u)}^{\alpha}$ (resp. $G_{(q,k,u)}^{\beta}$),
is a punctured code of $S_{(q,k)}^{\alpha}$ (resp. $S_{(q,k)}^{\beta}$),
and is the MacDonald code of type $\alpha$ (resp. $\beta$).
The MacDonald code $\mathcal{M}_{(q,k,u)}^{\alpha}$ is a code over $R_{q}$ of length $2^{2^q k}-2^{2^q u}$
and 2-dimension $2^{q}k$.
The MacDonald code $\mathcal{M}_{(q,k,u)}^{\beta}$ is a code over $R_{q}$ of length
$2^{\left( 2^q-1\right) (k-1)}(2^{k}-1)-2^{\left( 2^q-1\right)(u-1)}(2^{u}-1)$ and 2-dimension $2^{q}k$.

For example, if $q=2$, $k=3$ and $1 \leq u \leq 2$, there are two MacDonald codes of type $\alpha$
($\mathcal{M}_{(2,3,1)}^{\alpha}$ and $\mathcal{M}_{(2,3,2)}^{\alpha}$),
and two MacDonald codes of type $\beta$
($\mathcal{M}_{(2,3,1)}^{\beta}$ and $\mathcal{M}_{(2,3,2)}^{\beta}$).
If $\mathcal{U}_{\{1,2\}}=1+u_{1}+u_{2}+u_{1}u_{2}$ and $\mathcal{V}_{\{1,2\}}=u_{1}+u_{2}+u_{1}u_{2}$,
then the generator matrices of these codes are given by
\[
G_{(2,3,1)}^{\alpha }=\left(
\begin{array}{l|l|l|l}
\overset{256}{\overbrace{1\cdots 1}} & \overset{256}{\overbrace{
u_{1}\cdots u_{1}}} & \cdots  & \overset{256}{\overbrace{\mathcal{U}_{\{1,2\}}\cdots \mathcal{U}_{\{1,2\}}}} \\
\hline
G_{(2,2)}^{\alpha } & G_{(2,2)}^{\alpha } & \cdots  &
G_{(2,2)}^{\alpha }
\end{array}
\right),
\]
\[
G_{(2,3,2)}^{\alpha }=\left(
\begin{array}{l|l|l|l}
\overset{240}{\overbrace{0\cdots 0}} & \overset{256}{\overbrace{1\cdots 1}
} & \cdots  & \overset{256}{\overbrace{\mathcal{U}_{\{1,2\}}\cdots \mathcal{U}_{\{1,2\}}}} \\ \hline
\overset{16}{\overbrace{1\cdots 1}}\cdots \overset{16}{\overbrace{\mathcal{U}_{\{1,2\}}\cdots \mathcal{U}_{\{1,2\}}}
} & \overset{16}{\overbrace{0\cdots 0}}\cdots \overset{16}{\overbrace{
\mathcal{U}_{\{1,2\}}\cdots \mathcal{U}_{\{1,2\}}}} & \cdots  & \overset{16}{\overbrace{0\cdots 0}}\cdots
\overset{16}{\overbrace{\mathcal{U}_{\{1,2\}}\cdots \mathcal{U}_{\{1,2\}}}} \\ \hline
G_{(2,1)}^{\alpha }\setminus\overset{16}{\overbrace{01\cdots \mathcal{U}_{\{1,2\}}}}  & G_{(2,1)}^{\alpha } & \cdots  &
G_{(2,1)}^{\alpha }
\end{array}
\right)
\]

\[
G_{(2,3,1)}^{\beta }=\left(
\begin{array}{l|l|l|l}
\overset{256}{\overbrace{1\cdots 1}} & \overset{23}{\overbrace{0\cdots 0}}
 & \cdots  & \overset{24}{\overbrace{\mathcal{V}_{\{1,2\}}\cdots \mathcal{V}_{\{1,2\}}}} \\ \hline
\overset{16}{\overbrace{0\cdots 0}}\cdots \overset{16}{\overbrace{\mathcal{U}_{\{1,2\}}\cdots \mathcal{U}_{\{1,2\}}}
} & \overset{16}{\overbrace{1\cdots 1}}\overset{7}{\overbrace{u_{1}\cdots \mathcal{V}_{\{1,2\}}
}} & \cdots  & \overset{16}{\overbrace{1\cdots 1}}\overset{8}{\overbrace{
0u_{1}\cdots \mathcal{V}_{\{1,2\}}}} \\ \hline
\overset{16}{\overbrace{G_{(2,1)}^{\alpha }}}\cdots \overset{16}{\overbrace{
G_{(2,1)}^{\alpha }}} & \overset{16}{\overbrace{G_{(2,1)}^{\alpha }}}
\overset{7}{\overbrace{1\cdots 1}} & \cdots  & \overset{16}{\overbrace{
G_{(2,1)}^{\alpha }}}\overset{8}{\overbrace{1\cdots 1}}
\end{array}
\right)
\]

\[
G_{(2,3,2)}^{\beta }=\left(
\begin{array}{l|l|l|l}
\overset{256}{\overbrace{1\cdots 1}} & \overset{24}{\overbrace{
u_{1}\cdots u_{1}}} & \cdots  & \overset{24}{\overbrace{\mathcal{V}_{\{1,2\}}\cdots \mathcal{V}_{\{1,2\}}}} \\
\hline
G_{(2,2)}^{\alpha } & G_{(2,2)}^{\beta } & \cdots  & G_{(2,2)}^{\beta
}
\end{array}
\right)
\]

Using the previous notation, we have the following results.
\begin{thm}
\label{6}
Let $\mathcal{M}_{(q,k,u)}^{\alpha}$ and $\mathcal{M}_{(q,k,u)}^{\beta }$
be the MacDonald codes of types $\alpha $ and $\beta$, respectively, over $R_{q}$.
Then $\Gamma_{q}(\mathcal{M}_{(q,k,u)}^{\alpha})$ and $\Gamma_{q}(\mathcal{M}_{(q,k,u)}^{\beta })$ are the concatenation of
$2^{2^{q-1}k}$  MacDonald codes of types $\alpha $ and $\beta$,
respectively, over $R_{q-1}$.
\end{thm}
\pf
The proof is similar to those for Theorems \ref{thm:11} and \ref{thm:5}.
\qed

\begin{thm}
If $\mathcal{M}_{(q,k,u)}^{\alpha}$ is the MacDonald code of type $\alpha$ over $R_{q}$,
then
\[\Gamma_{q}\left( \Gamma_{q-1}\cdots \left(\Gamma_{2}\left(
\mathcal{M}_{(2,k,u)}^{\alpha}\right) \right) \right)
=\left( \mathcal{M}_{(1,k,u)}^{\alpha}\mathcal{M}_{(1,k,u)}^{\alpha}\cdots \mathcal{M}_{(1,k,u)}^{\alpha} \right)\]
is the concatenation of $2^{2^{q\left(\frac{q-1}{2} \right)k }}$
$\mathcal{M}_{(1,k,u)}^{\alpha}$ codes,
where $\mathcal{M}_{(1,k,u)}^{\alpha}$ is the MacDonald code of type $\alpha $ over $R_{1}$).

If $\mathcal{M}_{(q,k,u)}^{\beta}$ is the MacDonald code of type $\beta$ over $R_{q}$, then
\[
\left( \Gamma _{q}\left( \Gamma _{q-1}\cdots \left( \Gamma_{2}\left( \mathcal{M}_{(2,k,u)}^{\beta }\right) \right) \right)
=\left( \mathcal{M}_{(1,k,u)}^{\beta }\mathcal{M}_{(1,k,u)}^{\beta }\cdots \mathcal{M}_{(1,k,u)}^{\beta }\right) \right),
\]
is the concatenation of $2^{2^{q\left(\frac{q-1}{2} \right)k}}$ copies of $\mathcal{M}_{(1,k,u)}^{\beta}$,
where $\mathcal{M}_{(1,k,u)}^{\beta }$ is the MacDonald code of type $\beta $ over $R_{1}$.
\end{thm}
\pf
The proof is similar to those for Theorems \ref{thm:3} and \ref{thm:12}.
\qed

In the remainder of this paper,
we denote by $\mathcal{M}_{T,\alpha}$ and $\mathcal{M}_{T,\beta}$ the torsion codes of $\mathcal{M}_{(q,k,u)}^{\alpha}$ and $\mathcal{M}_{(q,k,u)}^{\beta}$, respectively.
Next, the Hamming weight distributions of $\mathcal{M}_{T,\alpha}$ and $\mathcal{M}_{T,\beta}$ are obtained.

\begin{thm}
\label{22}
The torsion code $\mathcal{M}_{T,\alpha}$ is a linear code with parameters
$(2^{2^{q}k}-2^{2^{q}u};k;2^{2^{q}k-1}-2^{2^{q}u-1})$.
The number of codewords with Hamming weight $2^{2^{q}k-1}-2^{2^{q}u-1}$ is equal to $2^{k}-2^{k-u}$,
the number of codewords with Hamming weight $2^{2^{q}k-1}$ is equal to $2^{k-u}-1$, and there is one codeword of zero weight.
\end{thm}

\pf
The generator matrix of the torsion code $\mathcal{M}_{T,\alpha}$ is obtained by replacing $u_{1}u_{2}\cdots u_{q}$ by $1$
in the matrix $u_{1}u_{2}\cdots u_{q}G_{(q,k,u)}^{\alpha}$.
Similar to the proof of \cite[Lemma 3.1]{{alashkar}}, the proof is by induction on $k$ and $u$.
It is clear that the result holds for $k = 2$ and $u = 1$.
Suppose the result holds for $k - 1$ and $1 \leq u \leq k-2$.
Then for $k$ and $1 \leq u \leq k-1$,
the matrix $u_{1}u_{2}\cdots u_{q}G_{(q,k,u)}^{\alpha}$ has the form
\begin{equation}
u_{1}u_{2}\cdots u_{q}G_{(q,k,u)}^{\alpha}=\left(
\begin{array}{ccc}
u_{1}u_{2}\cdots u_{q}G_{(q,k)}^{\alpha} & \left\backslash {}\right. & \frac{0_{2^{2^{u}}\times
\left( k-u\right) }}{u_{1}u_{2}\cdots u_{q}G_{(q,u)}^{\alpha }}
\end{array}
\right).
\tag{10}
\end{equation}
Then each nonzero codeword of $u_{1}u_{2}\cdots u_{q}G_{(q,k,u)}^{\alpha}$ has Hamming weight
$2^{2^{q}k-1}-2^{2^{q}u-1}$ or $2^{2^{q}k-1}$,
and the dimension of the torsion code $\mathcal{M}_{T,\alpha}$ is $k$.
Hence, the number of codewords with Hamming weight $2^{2^{q}k-1}-2^{2^{q}u-1}$
is $2^{k}-2^{k-u}$, and the number of codewords with Hamming weight $2^{2^{q}k-1}$ is $2^{k-u}-1$.
\qed
\begin{thm}
The Hamming, Lee and homogeneous weight distributions of $\mathcal{M}_{(q,k,u)}^{\alpha}$ are
\begin{enumerate}
\item[(i)]
$A_{Ham}(0)=1$, $A_{Ham}(2^{2^{q}k-1}-2^{2^{q}u-1})=2^{k}-2^{k-u}$, and $A_{Ham}(2^{2^{q}k-1})=2^{k-u}-1$.
\item[(ii)]
$A_{Lee}(0)=1$, $A_{Lee}(2^{2^{q}k+1})=2^{2^{q}(k-u)}-1$, and $A_{Lee}(2^{2^{q}k+1}-2^{2^{q}u+1})=2^{2^{q}(k-u)}(2^{2^{q}u}-1)$.
\item[(iii)]
$A_{hom}(0)=1$, $A_{hom}(2^{2^{q}k+1})=2^{2^{q}(k-u)}-1$, and $A_{hom}(2^{2^{q}k+1}-2^{2^{q}u+1})=2^{2^{q}(k-u)}(2^{2^{q}u}-1)$.
\end{enumerate}
\end{thm}

\pf
By Lemma \ref{lem:3} and (\ref{equation:44}), there are codewords of $\mathcal{M}_{(q,k,u)}^{\alpha}$ with Hamming weight
$2^{2^{q}k-1}-2^{2^{q}u-1}$ or $2^{2^{q}k-1}$, and
Lee and homogeneous weights $2^{2^{q}k+1}$ or $2^{2^{q}k+1}-2^{2^{q}u+1}$.
Furthermore, by Theorem \ref{22} the dimension of the torsion code $\mathcal{M}_{T,\alpha}$ is $k$.
Thus we have $2^{k-u}-1$ codewords of Hamming weight $2^{2^{q}k-1}$ and $2^{2^{q}k-1}-2^{2^{q}u-1}$ codewords
of Hamming weight $2^{k}-2^{k-u}$.
\qed

\begin{thm}
The torsion code $\mathcal{M}_{T,\beta}$ is a linear code with parameters
$(2^{\left( 2^q-1\right) (k-1)}(2^{k}-1)-2^{\left( 2^q-1\right) (u-1)}(2^{u}-1);k;2^{2^{q}k-2^{q}}-2^{2^{q}u-2^{q}})$.
The number of codewords with Hamming weight $2^{2^{q}k-2^{q}}-2^{2^{q}u-2^{q}}$
is $2^{k}-2^{k-u}$, the number of codewords with Hamming weight $2^{2^{q}k-2^{q}}$ is $2^{k-u}-1$,
and there is one codeword of weight $0$.
\end{thm}

\pf
The proof is similar to that for Theorem \ref{22}.
\qed

\subsection{Binary Gray Images of MacDonald Codes of Types $\alpha$ and $\beta$ over $R_{q}$}

The binary Gray images of the MacDonald codes of types $\alpha$ and $\beta$ are considered in this section.

\subsubsection{Binary Gray Images of MacDonald Codes of Type $\alpha$}

We now determine the binary images of the MacDonald codes of type $\alpha$ over $R_{q}$.
The first theorem considers the Lee weight and the second theorem considers the homogeneous weight.
\begin{thm}
\label{thm:13}
Let $\mathcal{M}_{(q,k,u)}^{\alpha}$ be the MacDonald code of type $\alpha$ over $R_{q}$ of
length $2^{2^{q}k}-2^{2^{q}u}$, 2-dimension $2^{q}k$ and minimum Lee weight $d_{Lee}$.
Then $\Psi _{Lee}(S_{(q,k)}^{\alpha})$ is the concatenation of $\dfrac{2^{2^{q}k+q}-2^{2^{q}u+q}}{2^{k}-2^{u}}$
binary MacDonald codes with parameters
$[2^{2^{q}k+q}-2^{2^{q}u+q};k;d_{Ham}=2^{2^{q}k+q-1}-2^{2^{q}u+q-1}]$.
\end{thm}
\pf
The proof is similar to that of Theorem \ref{thm:4}.
\qed
\begin{thm}
Let $\mathcal{M}_{(q,k,u)}^{\alpha}$, be the MacDonald code of type $\alpha$ over $R_{q}$ of
length $2^{2^{q}k}-2^{2^{q}u}$, 2-dimension $2^{q}k$ and minimum homogeneous weight $d_{hom}$.
Then $\Psi _{hom}(S_{(q,k)}^{\alpha})$ is the concatenation of $\dfrac{2^{2^{q}k+q+1}-2^{2^{q}u+q+1}}{2^{k}-2^{u}}$
binary MacDonald codes with parameters
$[2^{2^{q}k+q+1}-2^{2^{q}u+q+1};k;d_{Ham}=2^{2^{q}k+q}-2^{2^{q}u+q}]$.
\end{thm}
\pf
The proof is similar to that of Theorem \ref{thm:4}.
\qed
\subsubsection{Binary Gray Images of MacDonald Codes of Types $\beta$}
The binary Gray images of the MacDonald codes of type $\beta$ are now given.

\begin{thm}
Let $\mathcal{M}_{(q,k,u)}^{\beta}$
be the MacDonald code of type $\beta$ over $R_{q}$ of
length $2^{\left( 2^q-1\right) (k-1)}(2^{k}-1)-2^{\left( 2^q-1\right) (u-1)}(2^{u}-1)$,
2-dimension $2^{q}k$ and minimum Lee weight $d_{Lee}$.
Then $\Psi _{Lee}(S_{(q,k)}^{\beta})$ is the concatenation of $\dfrac{2^{\left( 2^q-1\right) (k-1)+q}(2^{k}-1)-2^{\left( 2^q-1\right) (u-1)+q}(2^{u}-1)}{2^{k}-2^{u}}$ copies of the binary MacDonald code with parameters
$
[2^{( 2^q-1) (k-1)+q}(2^{k}-1)-2^{( 2^q-1) (u-1)+q}(2^{u}-1);k;d_{Ham}=2^{( 2^q-1) (k-1)+q-1}(2^{k}-1)-2^{( 2^q-1) (u-1)+q-1}(2^{u}-1)].
$
\end{thm}
\pf
The proof is similar to that of Theorem \ref{thm:4}.
\qed

\begin{thm}
Let $\mathcal{M}_{(q,k,u)}^{\beta}$ be the MacDonald code of type $\beta$ over $R_{q}$ of length
$2^{\left( 2^q-1\right) (k-1)}(2^{k}-1)-2^{\left( 2^q-1\right) (u-1)}(2^{u}-1)$),
2-dimension $2^{q}k$ and minimum homogeneous weight $d_{hom}$.
Then $\Psi_{hom}(S_{(q,k)}^{\beta})$)
is the concatenation of $\dfrac{2^{( 2^q-1) (k-1)+(q+1)}(2^{k}-1)-2^{( 2^q-1) (u-1)+(q+1)}(2^{u}-1)}{2^{k}-2^{u}}$
binary MacDonald codes with parameters
$
[2^{( 2^q-1) (k-1)+(q+1)}(2^{k}-1)-2^{( 2^q-1) (u-1)+(q+1)}(2^{u}-1);k;d_{Ham}=2^{( 2^q-1) (k-1)+q}(2^{k}-1)-2^{( 2^q-1) (u-1)+q}(2^{u}-1)].
$
\end{thm}

\pf
The proofs are similar to those for Theorems \ref{thm:15} and \ref{thm:16}.
\qed

\section{The Repetition Codes over $R_{q}$ and their Covering Radius}

The repetition code $C$ over a finite field $\mathbb{F}_{q}$ is an $[n; 1; n]$ linear code.
The covering radius of $C$ is $\lfloor \frac{n(q-1)}{q}\rfloor$ \cite{Gupta}.
We begin by defining the repetition codes over ${R}_{q}$.
Let
\[
\mathcal{U}_{A}=\left( 1+\sum\limits_{\substack{ A\subseteq \left\{ 1,2,\ldots,q\right\}  \\
c_{\varnothing}=0\vee A\neq \varnothing }}c_{A}u_{A}\right),
 \]
and
\[
\mathcal{V}_{A}=\sum\limits_{\substack{ A\subseteq \left\{ 1,2,\ldots,q\right\}  \\
c_{\varnothing}=0\vee A\neq \varnothing }}c_{A}u_{A}.
\]
Two types of repetition codes can be defined over ${R}_{q}$.

\textbf{Type 1} The repetition codes $C_{c}$ generated by
\[
G_{c}=\left(\overset{n}{\overbrace{cc\cdots c}} \right),
\]
where $c$ is an element of $R_{q}$-$\{0,u_{1}u_{2} \cdots u_{q}\}$.

\textbf{Type 2} The repetition codes $C_{u_{1}u_{2} \cdots u_{q}}$ generated by
\[
G_{u_{1}u_{2} \cdots u_{q}}=\left(\overset{n}{\overbrace{u_{1}u_{2} \cdots u_{q}u_{1}u_{2} \cdots u_{q}\cdots u_{1}u_{2} \cdots u_{q}}} \right).
\]

\begin{thm}
\label{thm:35}
The covering radius of the repetition codes over $R_{q}$ is given by
\begin{enumerate}
\item[(i)] $r_{hom}(C_{c})=2^{q}n$ and $r_{Lee}(C_{c})=2^{q}n$.
\item[(ii)] $r_{hom}(C_{u_{1}u_{2} \cdots u_{q}})=2^{q+1}n$ and $r_{Lee}(C_{u_{1}u_{2} \cdots u_{q}})=2^{q}n$.
\end{enumerate}
\end{thm}

\pf
For part (i), by definition $r_{hom}(C_{c})$=$\max_{x\in (R_{q})^{n}}d \{x,C_{c}\}$.
Let
$x \in (R_{q}-\{0,u_{1}u_{2} \cdots u_{q}\})^{n}$.
Then as a direct consequence, for all $y \in C_{c}$ we have $d \{x,y\}=2^{q}n$, so that $r_{hom}(C_{c})=2^{q}n$.
By Proposition \ref{prop:33}, we obtain that $r_{Lee}(C_{c})=r_{Ham}(\Psi_{Lee}(C_{c}))=2^{q}n$.
The proof of part (ii) is similar.
\qed

Let $C$ be the linear code over $R_{q}$ generated by the matrix
\[
G=\left(\overset{n}{\overbrace{11\cdots 1}} \overset{n}{\overbrace{u_{1}u_{1}\cdots u_{1}}}\cdots\overset{n}{\overbrace{\mathcal{U}_{A}\mathcal{U}_{A}\cdots \mathcal{U}_{A}}}  \right).
\]
Then $C$ is the repetition code of length $(2^{2^{q}}-1)n$.

\begin{thm}
\label{thm:37}
A linear code $C$ generated by the matrix
\[G=\left(\overset{n}{\overbrace{11\cdots 1}} \overset{n}{\overbrace{u_{1}u_{1}\cdots u_{1}}}\cdots\overset{n}{\overbrace{\mathcal{U}_{A}\mathcal{U}_{A}\cdots \mathcal{U}_{A}}}  \right),
\]
has covering radius given by
\[
r_{hom}(C)=2^{2^{q}+q}n \mbox{ and }r_{Lee}(C)=(2^{2^{q}}-1)2^{q-1}n.
\]
\end{thm}

\pf
The vectors of $C$ generated by $G$ can be divided into three classes.
\begin{itemize}
\item[(1)] The vectors of $C$ with components from all the element of $R_{q}$
\[
x_{a}=(x_{1}x_{2}\cdots x_{n})\in C, x_{i} \in R_{q} \mbox{ for all } 1 \leq i \leq n.
\]
\item[(2)] The vectors of $C$ with components that are zero divisors of $R_{q}$
\[
x_{b}=(x_{1}x_{2}\cdots x_{n})\in C,  x_{i} \in \mathfrak{D}(R_{q}) \mbox{ for all }  1 \leq i \leq n.
\]
\item[(3)] The vectors of $C$ with components $0$ or $u_{1}u_{2}\cdots u_{q}$
\[
x_{c}=(x_{1}x_{2}\cdots x_{n})\in C, x_{i} \in \{0,u_{1}u_{2}\cdots u_{q}\} \mbox{ for all } 1 \leq i \leq n.
\]
\end{itemize}

For $x \in (R_{q})^{n}$, we have that $d(x,x_{a})=d(x,x_{b})=d(x,x_{c})=2^{2^{q}+q}n$,
so $r_{hom}(C)\geqslant 2^{2^{q}+q}n$.
On the other hand, for class (1), if $x=(11 \cdots 1) \in (R_{q})^{n}$ and
$x_{a}=(1u_{1} \cdots \mathcal{U}_{A}) \in (R_{q})^{n}$, then $x+x_{a}=(0(1+u_{1}) \cdots \mathcal{V}_{A})$
is a permutation equivalent to $x_{a}$ so that
\[
x+x_{a}=\sigma(x_{a}).
\]
Then $d(x,x_{a})\leq 2^{2^{q}+q}n$, and hence $r_{hom}(C)\leq 2^{2^{q}+q}n$.
For class (2), if $x=(11 \cdots 1) \in (R_{q})^{n}$ and $x_{b}=(u_{1}u_{2} \cdots \mathcal{V}_{A}) \in (\mathfrak{D}(R_{q}))^{n}$, so that
\[
 x+x_{b}=\left( (1+u_{1})(1+u_{2}) \cdots \mathcal{U}_{A}\right) \in (\mathfrak{U}(R_{q}))^{n}.
\]
Then $d(x,x_{b})\leq 2^{2^{q}+q}n$, and hence $r_{hom}(C)\leq 2^{2^{q}+q}n$.
For class (3), if $x=(11 \cdots 1) \in (R_{q})^{n}$ and $x_{c}=\left( 0(u_{1}u_{2}\cdots u_{q}) \cdots (u_{1}u_{2}\cdots u_{q})\right) \in (\mathfrak{D}(R_{q}))^{n}$, then
\[
x+x_{c}=\left( 1(1+u_{1}u_{2}\cdots u_{q}) \cdots (1+u_{1}u_{2}\cdots u_{q})\right) \in (\mathfrak{U}(R_{q}))^{n},
\]
so that $d(x,x_{c})\leq 2^{2^{q}+q}n$ and hence $r_{hom}(C)\leq 2^{2^{q}+q}n$.

By Proposition \ref{prop:33} we then have that
$r_{Lee}(C)=r_{Ham}(\Psi_{Lee}(C))=(2^{2^{q}}-1)2^{q-1}n$.
\qed

\section{The Covering Radius of Simplex and MacDonald Codes of Types $\alpha$ and $\beta$ over $R_{q}$}

We now determine the covering radius of simplex and MacDonald codes of types $\alpha$ and $\beta$ over $R_q$.
This requires the covering radius of the repetition code over $R_{q}$.

\subsection{The Covering Radius of Simplex Codes of Types $\alpha$ and $\beta$ over $R_{q}$}

The covering radius of simplex codes of types $\alpha$ and $\beta$ over $R_{q}$ is given by the following theorems.

\begin{thm}
\label{thm:36}
The covering radius of the simplex codes of type $\alpha$ over $R_{q}$ with respect to the homogeneous and Lee weights is
\begin{enumerate}
\item[(i)] $r_{hom}( S_{(q,k)}^{\alpha})=k\cdot2^{2^{q}k+q}$.
\item[(ii)] $r_{Lee}( S_{(q,k)}^{\alpha})=2^{(2^{q}+1)k+1}$.
\end{enumerate}
\end{thm}
\pf
For part (i), if $x \in (R_{q})^{n}$, we have $d_{hom}(x,S_{(q,k)}^{\alpha})=k\cdot2^{2^{q}k+q}$.
Hence by definition, $r_{hom}( S_{(q,k)}^{\alpha})\geqslant k\cdot2^{2^{q}k+q}$.
On the other hand, applying Proposition \ref{prop:34} and Theorem \ref{thm:37} gives
\[
\begin{array}{lll}
r_{hom}( S_{(q,k)}^{\alpha }) & \leq & r_{hom}\left(
\left[\overset{2^{2^{q}}(k-1)}{\overbrace{11\cdots 1}}\overset{2^{2^{q}}(k-1)
}{\overbrace{u_{1}u_{1}\cdots u_{1}}}\cdots \overset{2^{2^{q}}(k-1)}{\overbrace{
\mathcal{U}_{
A}\mathcal{U}_{A}\cdots \mathcal{U}_{A}}}\right]\right)  +2^{2^{q}}\cdot r_{hom}(
S_{(q,k-1)}^{\alpha }) \\
& \leq  & 2^{2^{q}k+q}+2^{2^{q}(k-1)+q}\cdot 2^{2^{q}}+\cdots +2^{q\cdot
2^{q}}\cdot r_{hom}( S_{(q,1)}^{\alpha }) \\
& \leq & 2^{2^{q}k+q}+2^{2^{q}(k-1)+q}\cdot 2^{2^{q}}+\cdots
+2^{2^{q}(k-q)+q}\cdot 2^{q\cdot 2^{q}} \\
& \leq & k\cdot 2^{2^{q}k+q}.
\end{array}
\]
For part (ii), from Proposition \ref{prop:33} we have
\[
r_{Lee}( S_{(q,k)}^{\alpha })=r_{Ham}( \Psi_{Lee}(S_{(q,k)}^{\alpha }))=2^{(2^{q}+1)k+1}.
\]
\qed

\begin{thm}
\label{thm:38}
The covering radius of the simplex codes of type $\beta$ over $R_{q}$ with respect the homogeneous and Lee weights is
\begin{enumerate}
\item[(i)] $r_{hom}( S_{(q,k)}^{\beta})=2^{2^{q}(k-2)+q}\left[ 2^{2^{q}}(k-2^{-q)})+4-2^{-q+1}\right] $.
\item[(ii)] $r_{Lee}( S_{(q,k)}^{\beta})=2^{(2^{q}-1)(k-1)+(q-1)}(2^{k}-1)$.
\end{enumerate}
\end{thm}
\pf
For part (i), if $x \in (R_{q})^{n}$, we have $d_{hom}(x,S_{(q,k)}^{\beta})=2^{2^{q}(k-2)+q}\left[ 2^{2^{q}}(k-2^{-q)})+4-2^{-q+1}\right]$.
Hence by definition, $r_{hom}( S_{(q,k)}^{\beta)}\geqslant 2^{2^{q}(k-2)+q}\left[ 2^{2^{q}}(k-2^{-q)})+4-2^{-q+1}\right]$.
On the other hand, applying Proposition \ref{prop:34} and Theorem \ref{thm:37} gives
\[
\begin{array}{lll}
r_{hom} (S_{(q,k)}^{\beta }) & \leq & r_{hom}\left( \left[
\overset{2^{2^{q}}(k-1)}{\overbrace{1\cdots 1}}\cdots \overset{
2^{(2^{q}-1)(k-1)}(2^{k}-1)}{\overbrace{\mathcal{V}_{A}\cdots
\mathcal{V}_{A}}}\right] \right) +r_{hom}( S_{(q,k-1)}^{\alpha })
+2^{2^{q}-1}\cdot r_{hom}( S_{(q,k-1)}^{\beta }) \\
& \leq & 2^{2^{q}(k-2)+q}\left( 2^{2^{q}-1}+2\right) +\cdots
+2^{2^{q}(k-2)+q}(k-1)+2^{q\cdot 2^{q}-q}\cdot r_{hom}(
S_{(q,2)}^{\beta }) \\
& \leq & 2^{2^{q}(k-2)+q}( 2^{2^{q}-1}+2) (
2-2^{-q})+\cdots +2^{2^{q}(k-2)+q}(k-1) \\
& \leq & 2^{2^{q}(k-2)+q}\left[ 2^{2^{q}}(k-2^{-q)})+4-2^{-q+1}\right].
\end{array}
\]
Then similar to the proof of part (ii) of Theorem \ref{thm:36}, the result follows.
\qed

\subsection{Covering Radius of MacDonald Codes of Types $\alpha$ and $\beta$ over $R_{q}$}

The covering radius of the MacDonald codes of types $\alpha$ and $\beta$ over $R_{q}$ is given by the following theorems.

\begin{thm}
\label{thm:39}
The covering radius of the MacDonald codes of type $\alpha$ over $R_{q}$ with respect to the homogeneous and Lee weights is
\begin{enumerate}
\item[(i)] For $u\leq e \leq k$, $r_{hom}(\mathcal{M}_{(q,k,u)}^{\alpha})\leq 2^{2^{q}k}-2^{2^{q}u}+r_{hom}(\mathcal{M}_{(q,e,u)}^{\alpha})$.
\item[(ii)] $r_{Lee}(\mathcal{M}_{(q,k,u)}^{\alpha})=2^{2^{q}k+(q-1)}-2^{2^{q}u+(q-1)}$.
\end{enumerate}
\end{thm}

\pf
For the first part, from Proposition \ref{prop:34} and Theorem \ref{thm:37},
if $u\leq e \leq k$, we have
\[
\begin{array}{lll}
r_{hom}(\mathcal{M}_{(q,k,u)}^{\alpha }) & \leq &
(2^{2^{q}}-1)(2^{2^{q}k-2^{q}})+r_{hom}(\mathcal{M}_{(q,k-1,u)}^{\alpha }) \\
& \leq & (2^{2^{q}}-1)(2^{2^{q}k-2^{q}})+(2^{2^{q}}-1)(2^{2^{q}k-(2^{q}-2)})+\cdots
+(2^{2^{q}}-1)2^{2^{q}e} \\
&  & +r_{hom}(\mathcal{M}_{(q,e,u)}^{\alpha }) \\
& \leq & 2^{2^{q}k}-2^{2^{q}e}+r_{hom}( \mathcal{M}_{(q,e,u)}^{\alpha}).
\end{array}
\]
For the second part, by Proposition \ref{prop:33}, we obtain that
\[r_{Lee}(\mathcal{M}_{(q,k,u)}^{\alpha })=r_{Ham}(\Psi_{Lee}(\mathcal{M}_{(q,k,u)}^{\alpha }))=2^{2^{q}k+(q-1)}-2^{2^{q}u+(q-1)}.
\]
\qed
\begin{thm}
The covering radius of the MacDonald codes of type $\beta$ over $R_{q}$ with respect to the homogeneous and Lee weights is
\begin{enumerate}
\item[(i)] For $u\leq e \leq k$, $r_{hom}( \mathcal{M}_{(q,k,u)}^{\beta})\leq 2^{\left( 2^q-1\right) (k-1)}(2^{k}-1)-2^{\left( 2^q-1\right)(u-1)}(2^{u}-1)+r_{hom}( \mathcal{M}_{(q,e,u)}^{\beta})$.
\item[(ii)] $r_{Lee}(\mathcal{M}_{(q,k,u)}^{\beta})=2^{\left( 2^q-1\right) (k-1)+(q-1)}(2^{k}-1)-2^{\left( 2^q-1\right)(u-1)+(q-1)}(2^{u}-1)$.
\end{enumerate}
\end{thm}
\pf
 For the first part, from
Proposition \ref{prop:34} and Theorem \ref{thm:37},
if $u\leq e \leq k$, we have
\[
\begin{array}{lll}
r_{hom}(\mathcal{M}_{(q,k,u)}^{\beta}) & \leq & (2^{2^{q}}-1)2^{\left( 2^q-1\right) (k-1)-\left( 2^q-1\right)}(2^{k}-1)+r_{hom}(\mathcal{M}_{(q,k-1,u)}^{\beta }) \\
& \leq & (2^{2^{q}}-1)2^{\left( 2^q-1\right) (k-1)-\left( 2^q-1\right)}(2^{k}-1)+(2^{2^{q}}-1)2^{\left( 2^q-1\right) (k-1)-(\left( 2^q-1\right)-2)}(2^{k}-1)\\
& +& \cdots +(2^{2^{q}}-1)2^{\left( 2^q-1\right) (k-1)-\left( 2^e-1\right)}(2^{k}-1) +r_{hom}(\mathcal{M}_{(q,e,u)}^{\beta }) \\
& \leq &2^{\left( 2^q-1\right) (k-1)}(2^{k}-1)-2^{\left( 2^q-1\right)(e-1)}(2^{e}-1)+r_{hom}( \mathcal{M}_{(q,e,u)}^{\beta}).
\end{array}
\]
For the second part,
By Proposition \ref{prop:33}, we obtain that
\[r_{Lee}(\mathcal{M}_{(q,k,u)}^{\beta})=r_{Ham}(\Psi_{Lee}(\mathcal{M}_{(q,k,u)}^{\beta}))=2^{\left( 2^q-1\right) (k-1)+(q-1)}(2^{k}-1)-2^{\left( 2^q-1\right)(u-1)+(q-1)}(2^{u}-1).
\]
\qed

\end{document}